\documentclass[12pt]{article}%
\usepackage{epsfig}
\usepackage{setspace}
\usepackage{graphicx}
\usepackage{multirow}
\usepackage{subfigure}
\usepackage{amsmath, amsthm, amssymb}
\usepackage{amsfonts}
\usepackage{enumerate}
\usepackage{booktabs}
\usepackage{natbib}
\usepackage{bm}
\usepackage{hyperref}
\usepackage{fancyhdr}
\usepackage{chngpage}
\usepackage{authblk}
\usepackage{pdflscape}
\usepackage{color}
\usepackage{amsmath}
\usepackage{amssymb}%
\usepackage{setspace}
\usepackage{cancel}
\usepackage[normalem]{ulem}
\providecommand{\U}[1]{\protect\rule{.1in}{.1in}}
\newcommand{\bX}{\bm{X}}

\newcommand{\bg}{\bm{g}}

\newcommand{\bx}{\bm{x}}

\newcommand{\bY}{\mathbf{Y}}

\newcommand{\bv}{\mathbf{v}}

\newcommand{\bZ}{\mathbf{Z}}

\newcommand{\bU}{\mathbf{U}}

\newcommand{\bu}{\mathbf{u}}

\newcommand{\bbeta}{{\bm{\beta}}}
\newcommand{\bomega}{{\bm{\omega}}}

\newcommand{\bmu}{{\bm{\mu}}}

\begin{document}
	\linespread{1.5}
	\title{\textbf{Selection of the Optimal Personalized Treatment  from Multiple Treatments with  Right-censored Multivariate Outcome Measures}}
	\author{}
	\author{Chathura Siriwardhana$^{1}$,  K.B. Kulasekera$^{2}$,  and  Somnath Datta$^{3}$
		
		\emph{\small $^1$Department of Quantitative Health Sciences,
			University of Hawaii John A. Burns School of Medicine, HI, USA,}
		\emph{\small $^2$Department of Bioinformatics \& Biostatistics,University of Louisville, Louisville, KY, USA.}
		\emph{\small $^3$Department of Biostatistics, University of Florida, Gainesville, FL, USA,}
	}
	\date{}
	\maketitle

	\begin{abstract}
		We propose a novel personalized concept for the optimal treatment selection for a situation where the response is a multivariate vector, that could contain right-censored variables such as survival time. The proposed method can be applied with any number of treatments and outcome variables, under a broad set of models. Following a working semiparametric Single Index Model that relates covariates and responses, we first define a patient-specific composite score, constructed from individual covariates. We then estimate conditional means of each response, given the patient score, correspond to each treatment, using a nonparametric smooth estimator. Next, a rank aggregation technique is applied to estimate an ordering of treatments based on ranked lists of treatment performance measures given by conditional means. We handle the right-censored data by incorporating the inverse probability of censoring weighting to the corresponding estimators. An empirical study illustrates the performance of the proposed method in finite sample problems. To show the applicability of the proposed procedure for real data, we also present a data analysis using HIV clinical trial data, that contained a right-censored survival event as one of the endpoints.   
		
		
	\end{abstract}
	
	\section{Introduction}
	
	The goal of personalized medicine is to use data to improve decision making in health care to provide the ``best" outcome for a patient based on his/her individualized features. The principle of personalized medicine is centuries old, but the idea advanced dramatically after the introduction of randomized controlled clinical trials. The primary aim in a majority of clinical trials is to make only a population-level decision but not an individualized decision that accounts for patient heterogeneity. But the increasing availability of data from such studies has increased the awareness of heterogeneity in both patient characteristics and outcomes and lead to new evidence-based medicine concepts. Over the last two decades, the statistical methodology in personalized medicine has led to new methodologies and insights, owing to advancement in computational power, bioinformatics discoveries, and access to electronic health data [1-5]. The basic premise of many existing personalized strategies is to make data driven decisions to optimize a targeted patient outcome while looking at the patient profile. In real-life situations, the success of treatment does not fully reflect through a single outcome as a variety of factors may compel both patients and clinicians to consider recovery in a rather broad view. For example, in the treatment for type 2 Diabetes, control in HbA1c, systolic blood pressure, low-density lipoproteins, cholesterol levels, and prevention from hypoglycemia and weight gain have been suggested as therapeutic goals to address the net clinical utility of a treatment [6]. In cancer studies, although the overall survival is considered the most important in this setting, a variety of other factors {such as} reduction in tumor size or eradication of cancerous cells, considered to be meaningful outcomes [7]. Similarly, in other situations where the disease is a life-threatening condition, time-to-event outcomes (e.g., overall survival) are commonly considered the best outcome, but there could be other factors with secondary importance, such as those relating to the quality of life and economic impact. Also, it is common to use a collection of surrogate outcomes during the early development of treatments [8].
	
	This work is a culmination of our previous contributions in the area of personalized medicine, more specifically, optimal treatment selection based on patient characteristics measured through covariates. We consider a situation where randomized trial data are available on patients where a number of responses (or a multi-dimensional response) were observed per patients. In addition, one or more of these responses are time-to-event and therefore subject to right censoring. This is the main novel feature of the methodology presented here and extends the papers by Siriwardhana et al. [5,9], where only completely observed responses were observed.
	
	This paper was motivated in part by a clinical trial data set on HIV patients which was only partially analyzed by the methodology available thus far in earlier articles by Siriwardhana et al. [5, 9]. Amongst other clinical characteristics such as the CD4 and CD8 counts, disease-free survival times were also observed which were subject to right censoring. However, the survival information was not directly implementable in the optimal treatment selection procedures because they were right-censored. We show in this work how to adjust the procedures in Siriwardhana et al. [9] to incorporate right-censored training data. The HIV data set is reanalyzed based on a trivariate response leading to better predictions of optimal treatment selection.
	
	The remainder of the paper is organized as follows: In Section 2, we discuss the proposed methodology. Section 3 includes simulation results, followed by real-life data illustration in Section 4. Finally, the main body of the paper ends with a discussion in Section 5.

	\section{Treatment Selection}
	
	{In this section, we describe the proposed personalized treatment selection procedure. Consider a situation where we observe {$J$} number of response variables for a patient undergoing a treatment selected from $K$ possible treatment options and, with no loss of generality, consider that larger values of individual responses are indicative of better outcomes. We assume that there is at least one response among $J$ responses is a survival type outcome that can be subjected to right censoring. Suppose {$\bY^{*}_{k}=(Y^{*}_{1k},...,Y^{*}_{Jk})'$} provides the $J$ dimensional vector of responses under the {$k$th} treatment option, for $r$ dimensional covariate vector $\bX$. In this work, we assume that we have data from a randomized clinical trial (RCT) study that provides responses and covariate information of a set of patients randomized into $K$ arms. It is important to note that, a dataset resulting from an RCT trial, does not have {$J\times K$} matrix of counterfactuals  $(\bY^{*}_1,...,\bY^{*}_K)$ for a single patient; hence, it is impossible to sample from the joint distribution of $(\bY^{*}_1,...,\bY^{*}_K,\bX)$ using such data.  Rather, we observe $K$ independent pairs of observations {$(\bY^{}_k,\bX_k)$} from  marginal distributions of {$(\bY^{*}_k,\bX)$ for $k=1,...,K$}, where {$\bY_k=(Y_{1k},...,Y_{Jk})'$}  (we refer readers to Siriwardhana et al. [5] for detailed discussion). 
		
		{As we describe later in this section,  we use the semi-parametric Single Index Model (SIM) as the working model for relating outcomes and covariates. This model provides a great flexibility in handling unknown nonlinearity between a response and a set of covariates. However, as SIM models could be different from actual mean models in real-life problems, we use the SIM model to obtain the first approximation of the conditional mean given a covariate. Then we produce a working model-based low dimensional score to use at the next level of smoothing via a fully nonparametric approach, to obtain an estimator for the conditional mean. Following Siriwardhana et al. [5,9], we use a patient’s covariate value $\bm{X}$ to obtain a lower dimensional composite patient score $U(\bX)$ that summarizes each patient’s characteristics.}
		
		{Following the ideas given in previous works, we use a patient’s covariate value $\bm{X}$ to obtain a lower dimensional composite patient score $U(\bX)$ that summarizes each patient’s characteristics.[5, 9, 10]}
		
		Here, we consider pairs of independent observations   $(\bY_k,\bX_k)$ from the marginal distribution of ${({\bY}^*_k,\bX)},  k=1,...,K$ to select the optimum treatment for $K$ treatments using the vectors of  smoothed conditional means for each treatment. We define   
		\begin{equation}\label{secondmu}\mu_{jk}(u_j)=E[Y_{jk}|U_{j}(\bm{X}_{k})=u_{j}];j=1,...,J;  k=1,...,K
		\end{equation}
		and  vectors  $\bmu_{k}(\bu)=(\mu_{1k}(u_{1}),...,\mu_{Jk}(u_{J}))'$  for $\bu=(u_1,...,u_J)'$ where components of these vectors correspond to each response.}

	{We rank the $K$ values for each component of {$\bmu_{k}(\bu)=(\mu_{1k}(u_{1}),...,\mu_{Jk}(u_{J}))'$}
		vectors {($k=1,...,K$)} to get size $ K$ vectors {$\bv_j(\bu)= (v_{j1}(\bu),...,v_{jK}(\bu))'$}, where {$v_{jk}(\bu)$} is the rank of {$\mu_{jk}$}  $k=1,...,K$ among {${\bm{\mu}^{j}(\bu)=}(\mu_{j1}(u_{j}),...,\mu_{jK}(u_{j}))'$} for each {$j$ (here $j=1,..,J$)} with the largest {$\mu_{jk}(u_j)$} value given the rank $1$. Then, we use an aggregation method given by Pihur et al. {[11, 12]} to combine these rank vectors to get an overall ranking of treatments $\bv^{*}=(v^{*}_{1},...v^{*}_{K})$. %Details of the rank aggregation steps are given in the appendix section of the article. 
		
		%In this article, we use the method proposed by Pihur et al. [10,11]  to aggregate these rank vectors for a given score vector {$\bU_0=(U_{1}(\bX_0),...,U_{J}(\bX_0))'$ corresponding to a covariate value $\bX_{0}$}. 
		%\textcolor{cyan}{***CHECK EQ NUMBERS**} 
		{For a set of weights {$w_j$, $j=1,...,J$}, selected appropriately, and for a distance measure $\gamma$ [10],  we minimize a quantity}
		{
			\begin{equation} 
			\label{distance}\psi(\bv)=\sum_{j=1}^{J}w_{j}\gamma(\bv, \bv_{j}(\bu)) 
			\end{equation}
		} 
		{over $ \bv \in P_K$, the set of all permutations of $\{1,...,K\}$. Among possible distance measures for $\gamma$, we used the weighted Spearman's Footrule distance [10], given by} 	
		\[
		{\sum_{l=1}^{I} |v_{l}-v_{r_l}| |M_{v_l}-M_{v_{r_l}}|^{\rho}},
		\]
		{where $ M_1,...,M_I$ is a list of real values, $r_1,...,r_I$ are the ranks of $M_l,l=1,...,I$  and $v_1,...,v_I$ is a permutation of integers $1,...,I$ and $\rho$ is a positive number.}
		
		{The above minimization yields a vector $\bv^{*}=(v^{*}_{1},...,v^{*}_{K})'$ where
			\begin{equation}\label{vstar}
			\bv^{*}=\mbox{arg}\min_{\bv \in P_{K}}\psi(\bv). 
			\end{equation}}
		
		In our approach, for {a covariate $\bX_0$ with a score $\bm{U_0}={U}(\bm{X_0})$}, we define the optimal treatment  as 
		\begin{equation}\label{k0}
		{k^{\ast}(\bm{U_{0}})=\mbox{arg}\ \min_{1\leq k\leq K}\{v^{*}_{k}\}}
		\end{equation}
		
		{Note that, in the unlikely event of observing multiple minimizers in \ref{vstar} or \ref{k0}, we randomly select one among those with ties.}
		
		{We link the $j$th component  $Y_{jk}$ of the response vector $\bY_k$ for the $k$th treatment and covariates $\bm{X}_{k}$ via a SIM model,
			\begin{equation}\label{model1}
			Y_{jk}=g_{jk}\left(  {\bm{\beta}}_{jk}^{\prime}\bm{X}_k\right)  +\epsilon_{jk}
			\end{equation}
			for $j=1,\ldots,J$ and $k=1,...,K$, where each $\bm{\beta}_{jk}$ is a $r$-vector of
			parameters, $g_{jk}$ is an unknown link function for which we assume some
			reasonable smoothness conditions to hold, and $\epsilon_{ik}$ are error terms with $E[\epsilon_{jk}|\bX]=0$. Furthermore we assume independence of $\epsilon_{jk}$s across $k=1,...,K$ for a fixed $j$ where these terms are correlated across $j$s for any given $k$. The SIM  formulation  provides flexibility and reasonable efficiency in modeling many types of data.
			
			As we described in the beginning of this section, we may observe right-censored outcomes for time-to-event type responses while using data sources such as historical clinical trial records. In our discussion below, we initially focus on explaining some aspects of the proposed strategy for completely observed responses, summarizing the procedure by Siriwardhana et al. [9].  Later, we generalize these ideas for problems with right-censored outcomes.}

		\subsubsection*{Treatment Selection with the Complete (uncensored) Data:}
		
		When complete response (i.e., uncensored) data available, observations are of the  form $(\bY_{ki}, \bX_{ki}) $ where $ \bY_{ki}=(Y_{1ki},...,Y_{Jki})'$ and $Y_{jki}$ indicates the $j$th component of the $i$th individual 
		under treatment $k$ with associated covariate values $\bm{X}_{ki},i=1,\ldots,n_{k}$. Then, for this data,
		relationship \eqref{model1} can be written as%
		{
			\begin{equation}
			Y_{jki}=g_{jk}({\bm{\beta}}_{jk}^{\prime}\bm{X}_{ki})+\epsilon_{jki}%
			,i=1,\ldots,n_{k}\text{.} \label{lm}%
			\end{equation}}
		Siriwardhana et al.[9] define a score vector $\bU(\bm{X})$ for a patient with $\bm{X}$ covariate as follows. First, we define,  
		\[
		S_{jk}\left(  \bm{X}\right)  =g_{jk}\left(  {\bm{\beta}}_{jk}^{\prime
		}\bm{X}\right)  -\max_{l\neq k}\left\{  g_{jl}\left(  {\bm{\beta}}_{jl}^{\prime
		}\bm{X}\right)  \right\}  .
		\]
		Next, define the $j$th components of the combined overall score  vectors as 
		{
			\begin{align}
			S_j\left(  \bm{X}\right)   &  =\max_{k}\left\{  S_{jk} \left( \bm{X} \right)    \right\} \nonumber\\
			\delta_j\left(  \bm{X}\right)   &  =\arg\max_{k}\left\{  S_{jk} \left( \bm{X} \right)   \right\}  .
			\end{align}
		}
		{The overall score is given as $\bU(\bm{X})=(U_1(\bm{X}),...,U_J(\bm{X}))'$  where $U_j(\bm{X})=(S_j(\bm{X}),\delta_j(\bm{X}))'$ for $j=1,...,J$.} {Note that, $\delta$ will be selected randomly among ties, in the improbable event that multiple treatments produce the largest $S_{jk}(\bm{X})$.}

		However, in reality, error distributions and model functions for models defined in (\ref{model1}) are unknown. Therefore standard function estimation method as described in Siriwardhana et al. [5,9] should be applied to estimate components of these score vectors. 
		
		In particular,  for any given vector $\bx$, let 
		\begin{align}\label{hats}
		{\hat{S}_{jk}(\bx)}& {= \hat{g}_{jk}\left( \hat{ {\bm{\beta}}}_{jk}^{\prime
			}\bm{x}\right)-\max_{l\neq k}\left\{  \hat{g}_{jl}\left( \hat{ {\bm{\beta}}}_{jl}^{\prime
			}\bx\right)  \right\}\nonumber}\\
		\hat{S}_{j}\left(  \bm{x}\right)   &  =\max_{k}\left\{  \hat{S}_{jk}(\bx)\right\} \nonumber\\
		\hat{\delta}_{j}\left(  \bm{x}\right)   &  =\arg\max_{k}\left\{  \hat{S}_{jk}(\bx)\right\} \nonumber \\
		&\mbox{and}\nonumber \\
		{\hat{U}_{j}(\bx)} &=  {(\hat{S}_{j}\left(  \bm{x}\right), \hat{\delta}_{j}\left(  \bm{x}\right))'; j=1,...,J.}
		\end{align}
		
		Siriwradhana et al. [9] proposed estimators for $\mu_{jk}(u_{j})$, $k=1,...,K$ at a given  $u_{j}=(s_{j},d_{j})^{\prime}$ as follows. Here  $\omega$ is a kernel function with $\omega \geq 0$ and $\int_{}^{}\omega(t)dt=1$,  and  {$h_{k}, k=1,...,K$} are a set of smoothing parameters. 
		{
			\begin{equation}
			\hat{\mu}_{jk}\left(  {u_{j}}\right)  =\frac{\sum_{i=1}^{n_k} Y_{jki}
				{\omega}%
				\left(  (s_{j}-\hat{S_{j}}(\bm{X}_{ki}))/{h_{jk}}\right)  I\left(  \hat{\delta}_{j}\left(  \bm{X}_{ki}\right)
				=d_{j}\right)  }{\sum_{i=1}^{n_k} 
				{\omega}%
				\left(  (s_{j}-{\hat{S_{j}}} (\bm{X}_{ki}))/{h_{jk}}\right)  I\left(  \hat{\delta}_{j}\left(  \bm{X}_{ki}\right)
				=d_{j}\right) }
			\label{temp_019}%
			\end{equation}}
		where $I(A)$ is the indicator of $A$. {The bandwidth selection for estimating $\mu_{jk}$s is a challenging issue and we do not investigate the optimal bandwidth selection for this problem. However, as Siriwardhana et al. [9] suggested, the method given in Wand and Jones {[13]} for kernel smoothing provides a reasonable solution for this estimation problem. We adopt this method for the current work.}

		For a realization $\bx_0$ of the covariate  $\bX$, if one could find the corresponding realizations of the scores,
		$u_{j0}=(S_{j}(\bx_0), \delta_{j}(\bx_0))'$,  this allows estimating $\mu_{jk}(u_{j0}) $   by $\hat{\mu}_{jk}(u_{j0})$. However, due to aforementioned reasons, one may only find an estimate $\hat{u}_{j0}=( {\hat{s}_{j0}},\hat{\delta}_{j0})'$ of  ${u}_{j0}$  using (\ref{hats}) above. Thus, in practice 
		one may use  $\hat{\mu}_{jk}(\hat{u}_{j0})$  as the  estimate of $\mu_{jk}(u_{j0})$  for $j=1,...,J; k=1,...,K$, with $\hat{\bu}_{0}=(\hat{u}_{10},...,\hat{u}_{J0})'$. 
		
		\subsubsection*{Treatment Selection with Right-censored Data:}
		
		{In this section, we extend the treatment selection concept by Siriwardhana et al. [9] problems that are involved with the right-censored survival type data. When complete (i.e., uncensored) data are available, we estimate each component of the overall score {$\bU(\bX)=(U_1(\bm{X}),...,U_J(\bm{X}))'$}  via a set of estimated single index models or partial linear models, given by {(\ref{model1})}. Subsequently, we obtain an estimator for $\mu_{jk}(\hat{u}_j)$, $j=1,.., J$, $k=1,..,K$, by {(\ref{temp_019})}. Suppose there exists at least one response among $J$ that subject to right-censoring. Let $j$th, $j=1,..., J$, be such a time-to-event type right-censored outcome. Since some $Y_{jki}$'s, $i=1,...,n_k$, are now unobservable, the direct application of the previous technique is no longer possible.} 
		
		{The use of data weighting schemes for the purpose of bias reduction is well known in the statistical literature. In statistical literature, data weighting schemes are popularly applied for bias reduction in estimates. A common strategy to handle censored observations in regression setting is to introduce a re-weighing scheme such as the {Inverse Probability Censoring Weighting (IPCW)} to the original estimator developed for complete data, in a way that the bias caused due to censoring fades away asymptotically. The basics of this concept {were} first introduced by Koul, Susarla, and Van Ryzin {[14]} for the randomly right censored data in linear regression and, later extended by many other authors for various problems with censoring. For examples, Satten and Datta {[15]} described estimating the marginal survival time in the presence of time dependent covariates, using a re-weighted Kaplan-Meier estimate based on the IPCW weights calculated from Aalen’s additive hazard model {[16, 17]}. Recently, Siriwardhana et al. {[18]} described estimating the binary choice single index model incorporating IPCW for censored data.  Following a similar approach, we introduce a reweighing scheme to estimators of the SIM model and $\mu_{jk}(u_j)$.
			
			Let  $C_{jki}$ be the right-censoring time correspond to the $j$th time-to-event response, $j=1...,J$, for the subject $i$, $i=1,...,n_k$, who treated from the $k$th treatment group, $k=1,...,K$, and let $Y_{jki}$ is the true response of the individual. Thus, we observe the censored response $T_{jki}=\min\{Y_{jki}, C_{jki}\}$ with the censoring indicator $\delta^{'}_{jki}=I[Y_{jki}\leq C_{jki}].\,$ We denote the survival function of the censoring distribution by ${K^{c}_{jki}}(t)=E\{I[C_{jki}>t]\},$ with hazard function $\lambda^{c}_{jki}(t)\,$and cumulative hazard function $\Lambda^{c}_{jki}(t)$, at a time point $t$. }
		
		\subsubsection*{IPCW Re-weighted Single Index Estimator:}
		{We use an IPCW re-weighted Ichimura et al. {[19]} SIM model to estimate the model {(\ref{model1})},  which allows one to to calculate the score {$\bm{\hat{U}_{j}(X)}=(\hat{S}_{j}(\bm{X)},\hat{\delta}_{j}(\bm{X}))$} with respect to a right-censored outcome $j$, $j=1,...,J$, by $k=1,...,K$. {[20]} The covariate response relationship for the $j$th outcome by $k$th treatment can be represented as
			{
				\[
				Y_{jki}=f_{jk}(\bm{\theta}_{jk}'{\bm{X}_{ki}})+\epsilon_{jki}, i=1,...,n_{k},
				\]
			}
			for some function $f_{jk}(.)$ and parameter vector {$\bm{\theta}_{jk}$ ($\bm{\theta}_{jk}\in \mathcal{R}^{r}$)},  where $\epsilon_{jki}$’s are independent errors with $E(\epsilon_{jki}|{\bm{X}_{ki}})=0$ and bounded common variance $\sigma$.
			For the purpose of identifiability, we replace $\bm{\theta}_{jk}$ by a unit vector, 
			\[\bbeta_{jk}=\bm{\theta}_{jk}||\bm{\theta}_{jk}||^{-1}.\]
			where $||.||$ is the Euclidean norm. Thus, an
			equivalent model can be written as,
			{
				\[
				Y_{jki}={g}_{jk}(\bbeta_{jki}'{\bm{X}_{ki}})+\epsilon_{jki},
				\]}
			which has the same form of model {(\ref{model1})}, where $g_{jk}(.)$ is an unknown univariate smooth link function. {For the notational convenience, we suppress the treatment indicator $k$ and the response indicator $j$ to describe the IPCW re-weighted SIM estimation procedure.} 
			
			Assuming all ${\bm{Y}}$'s are completely observed, Ichimura et al. {[19]} proposed an estimator to estimate the above SIM model, that estimate the unknown function ${g}(.)\,$ at point $\nu$, by leave-one-out cross validation method, omitting the pair of ${({Y_{i}, \bX_{i}})}$, 
			\[
			\hat{g}_{-i}(\nu|\bbeta)  =\frac{\sum_{l\neq
					i}Y_l\phi_{\tilde{h}}(\nu-\bbeta^{'}\bm{X}_l)}{\sum_{l\neq i}\phi_{\tilde{h}}(\nu-\bbeta^{'}\bm{X}_l)}.
			\]
			{where, $\tilde{h}$ is a$\,$smoothing parameter, $\phi_{\tilde{h}}(.)=\phi(./\tilde{h}),\,$ and $\phi(.)$, is a} fixed kernel function with $\phi(.) \geq 0$ and $\int_{}^{}\phi(t)dt=1$. Ichimura et al. {[19]} showed, estimates of  $\bbeta$ and $\tilde{h}$ can be achieved by simultaneously minimizing the following objective function with respect to $\bbeta$ and $\tilde{h}$. 
			\[\hat{S}(\bbeta,\,\tilde{h})=\sum_i\{Y_i-\hat{g}_{-i}(\bbeta'\bm{X}_{i}|{\bbeta})\}^2\]
			As the above SIM estimator is no longer valid when the data are subject to right-censoring, we suggest an alternative SIM estimator that is capable of handling right-censored data under survival-type outcomes. This new estimator is primarily based on  the method proposed by Ichimura et al. {[19]} but it's re-weighted by a IPCW weighting scheme. We define leave-one-out re-weighed estimator of $g(.)$ as, 
			
			{
				\[
				\hat{g}_{-i}(\nu|\bbeta)=
				\frac{
					\sum_{l\neq	i}
					\frac{\delta^{'}_l\,T_l}{K^c_{l_{-i}}(T_l-)}\phi_{\tilde{h}}(\nu-\bbeta^{'}\bm{X}_l)
				}
				{\sum_{l\neq i} { \frac{\delta^{'}_l}{K^c_{l_{-i}}(T_l-)}}  \phi_{\tilde{h}}(\nu-\bbeta^{'}\bm{X}_l)	}.
				\]}
			
			Here, $K^c(T-)$ is the survival probability of an individual not being censored just before time $T$ with $K^{c}_i(t)=\prod_{s\geq t} [1-\lambda^{c}_{i}(t|\bar{Z_i}(s))ds]$ and $\bar{Z_i}(t)$ is a generalized covariate defined for the $i$th individual, which we will explain in detail in the sequel. Note that $K^{c}_i(t)$ does not have a survival function interpretation unless $\bar{Z_i}(t)$ is formed with non-time-varying covariates. In reality, we replace $K^{c}_i(.)$ by its corresponding estimator $\hat{K}^{c}_i(.)$ and later we will introduce a flexible model to estimate $K^{c}_i(t)$ for any time $t$.} 
		
		We estimate $\bbeta$ and $\tilde{h}$ by minimizing the following weighted objective function denoted by $({\hat{S}}^{'}(\bbeta, \tilde{h}))$ simultaneously with respect to both $\bbeta$ and $\tilde{h}$.
		{
			\[
			{\hat{S}}^{'}(\bbeta,\,\tilde{h})
			=\sum_i\frac{\delta^{'}_i}{K^{c}_{i}(T_i-)}\{T_i-\hat{g}_{-i}(%
			\bbeta'\bm{X}_{i}|\bbeta)\}^2
			\]
		}
		For the estimator of $\bbeta$ ($\hat{\bbeta}$) and optimal $\tilde{h}$ ($\tilde{h}_0$), $g(.)$ function at a new point $\nu_{0}=\bbeta^{'}\bm{x}_{0}$, can be estimated as,
		{
			\begin{equation}
			\hat{g}(\nu_{0}|\hat\bbeta)=
			\frac{
				\sum_{i}\frac{\delta^{'}_i\,T_i}{K^{c}_{i}(T_i-)}\phi_{\tilde{h}_0}(\nu_{0}-\hat\bbeta^{'}\bm{X}_i)}{\sum_{i} { \frac{\delta^{'}_l}{K^c_{i}(T_i-)}}      \phi_{\tilde{h}_0}(\nu_{0}-\hat\bbeta^{'}\bm{X}_i)}.
			\label{SIM-RW}
			\end{equation}}} 
	
	\subsubsection*{IPCW re-weighted Estimator for $\bm{\mu_{jk}(u_j)}$:}  
	{As described before, the estimator given by {(\ref{temp_019})} is intend to use for estimating $\mu_{jk}(u_j)$, $k=1,..,K$, when the outcome $j$, $j=1,..,J$, is uncensored. Now, we provide an IPCW re-weighted smooth mean estimator for estimating $\mu_{jk}(u_j)$,  to handle the right-censored survival outcomes, where the estimator {(\ref{temp_019})} is no longer valid. The new re-weighted estimator was obtained similarly as the approach followed to adjust the SIM estimator for censored case. This is given by,
		
		{
			\begin{equation}
			\hat{\mu}_{jk}\left(  {{u}_j}\right)  = \frac{\sum_{i=1}^{n_k} \frac{\delta^{'}_{jki}}{\hat{K}^{c}_{jki}(T_{jki}-)}T_{jki}
				{\omega}%
				\left(  (s_j-\hat{S_{}}({\bm{X}_{ki}}))/h_{jk}\right)  I\left(  \hat{\delta}_{}\left(  \bm{X}_{ki}\right)
				=d_j\right)  }{   \sum_{i=1}^{n_k}  {\frac{\delta^{'}_{jki}}{\hat{K}^{c}_{jki}(T_{jki}-)}}
				{\omega}%
				\left(  (s_j-\hat{S_{}}({\bm{X}_{ki}}))/h_{jk}\right)  I\left(  \hat{\delta}_{}\left(  {\bm{X}_{ki}}\right)
				=d_j\right)   },
			\label{temp_021}%
			\end{equation}}
		
		{for a score value ${u}_{j}=(s_j,d_j)$} correspond to a covariate $x_0$. The estimation of weights can be performed using the Aalen's additive model as described in the next section. 
		
		Following the estimation of IPCW re-weighted SIM models for {a right-censored outcome} $Y_{jk}$ and covariate $\bm{X}_{k}$ pairs, we subsequently estimate scores 	
		$\hat{U}_{j}(\bX)=(\hat{S}_{j}(\bX),\hat{\delta}_{j}(\bX))$ at each covariate point $\bX$, including the covariate of new patient $\bx_0$. Finally, $\hat{\mu}_{jk}\left(  {\hat{u}_j}\right)$ is obtained using {{(\ref{temp_021})}.} 
		
		{Now, if we have a set of $J$ responses that contain some right-censored responses, we will be able to calculate the estimated score {$\hat{\bu}_{0}=(\hat{u}_{10},...,\hat{u}_{J0})'$} and its corresponding {$\hat{\bmu}_{k}(\hat{\bu}_{0})=(\hat{\mu}_{1k}(\hat{u}_{10}),...,\hat{\mu}_{Jk}(\hat{u}_{J0}))'$} for a given new patient covariate value $x_0$. The estimated best treatment $k^{*}(\bu)$is obtained following steps given in {(\ref{distance})} to {(\ref{vstar})}.
			
			\subsubsection*{Estimation of IPCW Weights:}
			{We use the Aalen's nonparametric additive model {[16, 17]} to calculate IPCW weights, which provides a flexible structure to estimate the censoring hazards by allowing covariates to be varied over  time. Let $\bZ_i(t)$ be a generalized covariate defined for individual $i$, $1\leq i \leq n_k$, at time $t$, which may contain both baseline and additional covariates (could be time varying) than covariates of primary interest $\bX$ that affect the censoring hazards. Suppose $\bar{\bZ}_i(t)=\sigma\{\bZ_i(s):0\leq s < t\}$ is the observed covariate history prior to $t$.  The additive censoring hazard of $i$th individual at time $t$ is given by the  linear form  
				$
				\lambda^{c}_i(t|\bar{\bZ}_i(t))=\sum_{m=0}^{M}\eta_m(t)W_{im}(t),
				$ 
				where, $W_{i0}(t)\equiv1$ and $W_{im}(t)=f_{m}(\bar{Z}_i(t)),m=1,...,M,$ are possibly time-dependent functions of the past history of the covariate process for subject $i$ and $\eta_m(t)$ are unknown regression functions that measure the effect of corresponding covariate functions on the censoring hazard. Define $W_i(t)=(W_{i1}(t),..,W_{ip}(t))$. Then the Aalen's estimator of cumulative censoring hazard for the $j$th individual is given by,
				\[
				\Lambda^c_{i}(t|\bar{Z}_i(t))=\int_{0}^{t}\hat{\lambda}^c_{i}(u|\bar{Z}_{i}(t))du=\sum_{i=1}^{n_k}I(T_i\leq t)(1-\delta_i)W_i(T_i)R^{-1}(T_i)W_i(T_i), 
				\]
				where
				$
				R(t)=\sum_{i=1}^{n_k}I(T_i\geq t)W_i(t)W^{'}_i(t). 
				$
				The estimated IPCW weight for $i$th individual can be expressed as
				$
				\hat{K}^c_i(t)=exp(-\hat{\Lambda}^c_{i}(t|\bar{Z}_i(t))),
				$
				where, 
				$
				\Lambda^c_{i}(t|\bar{\bZ}_i(t))=\int_{0}^{t}\lambda^{c}_i(u|\bar{\bZ}_i(t))du.$
			}
			
			\section{Empirical Studies}
			
			In this section we present a  simulation study that investigates the properties of the proposed procedure in finite sample cases. 
			
			In this investigation we primarily focused on the accuracy of treatment assignment of a new (test) observation using the proposed technique based on simulated samples for $K$ treatment groups and $J$ responses. We fixed one of the responses among $J$ set to be a right-censored time-to-event type response. The censoring time for the right-censored variable was generated under both random and covariate dependent settings with rates ranged between 25\% and 50\%. To give a relevance for a personalized treatment scenario, we selected our model sets such that each model in a set dominates other competing models for some combination of covariate values. In other words, none of considered models fully dominate other models within the whole covariate space, that makes the optimal treatment to be a function of patients' covariate profiles. We used two types of nonlinear model functions, given by Model Set-1 (Table \ref{SM-1}) and Mode Set-2 (Table \ref{SM-2}), for the generation of mean response per each $k=1,...,K$ and $j=1,...,J$. The kernel function for all smoothing was taken to be a Normal probability density function ($N(0,1)$).The bandwidth selection for smooth the mean estimators was by the algorithm given by  Wand and Jones {[13]}.
			
			Primary steps of the simulation study are given below:

			\begin{enumerate}
				
				\item {Covariate $\bm{X}$: Simulate $K$ independent {$r=5$ dimensional  multivariate random samples of size $n$, letting distribution of each component of $\bm{X}$ as $U(-1,1)$}}. Select $n$ from the set {$\{100, 200, 400\}$}.

				\item {For the functions $\bg_{j1}(.),...\bg_{jK}(.)$ and index vectors $\bbeta_1,...,\bbeta_K$  generate treatment responses from model (\ref{model1}) for $K$ groups.} 
				
				\item {For each {$k$, $k=1,...,K$}, generate $n$ errors from either a {$J$} dimensional multivariate normal distribution or a multivariate double exponential  distribution with zero mean and a correlation matrix with off-diagonal elements given by {$\rho$} and dispersion parameter $\sigma$. Select $\sigma$ from the set $\{0.3,0.5,1.0\}$. Use the R package {\em mvtnorm} {[21]}  for multivariate normal case and use the package {\em LaplacesDemon} {[22]}  for generating  double exponential random variables.}   
				
				\item Fix the first component of the response vector ($j=1$) to be a right-censored survival time type response and generate censoring times ($C$) as follows. For censoring rate 25\% and 50\% cases, select parameters from Table {\ref{cen_1}} and Table {\ref{cens_2}} for Model Set 1 and 2, respectively. 
				
				\begin{enumerate}
					
					\item Random censoring time : Use the Exponential distributions with a scale parameter $\zeta$.
					
					\item Covariate dependent censoring : Use the Exponential distribution with the scaler parameter given by the indicator function, 
					\[C
					\sim I(\rho^{'}X > w)\exp(\zeta_{1})+I(\rho^{'}X\leq w)\exp(\zeta_{2}),
					\]
					with $w=0$ and an arbitrarily selected vector $\rho=(0.7,0.3,0,0.5,-0.5)'$. 
					
				\end{enumerate}
				
				\item Estimate corresponding SIMs for $k=1,..,K$ and $j=1,..,J$, using observed response covariate pairs: Use the IPCW re-weighted SIM estimator for the right censored responses and use the regular SIM estimation for the completely observed cases $j\neq1$.

				\item Estimate individual scores $U_{j}(\bX)=(S_j(\bX),\delta_j(\bX))'$ at each covariate value $\bX$ for $j=1,..,J$, following (\ref{hats}). This gives the estimated overall score $\hat{U}(\bX)=\big(\hat{U}_{1}(\bX), ..., \hat{U}_{J}(\bX)\big)$ for $\bX$. 
				
				\item {Generate a new covariate value $\bm{x}_0$ using the parameters in Step 1.  Then estimate the score $\hat{u}(\bm{x}_0)=(\hat{u}_{10}, ..., \hat{u}_{J0})$. 
					
					\item Calculate {$\hat{\mu}_{jk}(\hat{u}_{j0})$ for $j=1,...,J; k=1,...,K$}: Use the IPCW re-weighted estimator (\ref{temp_021}) for the right censored response $j=1$ and the regular estimator (\ref{temp_019}) for complete cases $j\neq1$.
					
					\item Estimate the corresponding {$\hat{k}^{\ast}$}  for  weights vector $\bm{w}$.}

				\item {Generate $K$  response vectors {$\bm{y_{0k}}=(g_{1k}(\bm{\beta}_{1k}'\bx_0),$ $..., g_{Jk}(\bm{\beta}_{Jk}'\bx_0))'+\bf{\epsilon_{0k}}$}. Obtain $\bf{\epsilon_{0k}}$ using the same $J$ dimensional multivariate distribution as in Step-3.}

				\item {Obtain rank vectors {$\tilde{\bv}_{j} $, $j=1,...,J$,} for   each row of the  mean matrix $(\bm{y_{01}},...,\bm{y_{0K}})$, and minimize
					{\begin{equation} \label{distance2}\psi(\bv)=\sum_{j=1}^{J}w_{j}\gamma(\bv, \tilde{\bv}_{k}) \end{equation}}
					over $P_{K}$ for same weights {$\bomega$}  above to get the corresponding aggregated vector $({\hat{v}^{*}}_1,...,{\hat{v}^{*}}_K)'$ and 
					define the  treatment assignment to be correct if
					{ \[\hat{k}^{\ast}=\mbox{arg}\ \min_{1\leq k\leq K}\{{\hat{v}^{*}}_{k}\}\]}
					for the {$\hat{k}^{\ast}$} corresponding to  {$\hat{\mu}_{jk}$s}.}
				
				\item {Repeat steps 1-6  $1000$ times.} 
				
			\end{enumerate}
			
			We conducted our simulation study for $K=3$ with $J=3$ and $J=4$ cases by fixing one of the responses to be a right-censored response (i.e., $j=1$ in each case). Tables 6-9 summarize frequencies of accurate treatment selection in 1000 test cases. These results are stratified for two choices of weights; equal and unequal cases. In the unequal weight scenario, we assigned a large weight to the right-censored outcome, to closely examine the impact of censoring on the selection performance. As we found, especially when a large weight was assigned to the censored-response, the accuracy dropped at a high censoring rate, but continued to improve as the per group size $n$ increased. In general, a performance drop should be expected for the high censoring rate and small sample combination, as the IPCW weights correct the censoring bias in an asymptotic fashion. For a given censoring rate, comparing random and covariate dependent censoring types, there were no noticeable differences in the selection performance. As to be expected, the selection accuracy drops when the error distribution has high variability. 
			
			In conclusion, we observed reasonable performance by the proposed technique while considering the complexity of overall the treatment selection problem. For instance, in Model Set-2, we used highly nonlinear mean functions based on sine and cosine functions, which creates a complex treatment selection scenario, however, the selection accuracy remained reasonably high for those cases, showing the potential of the proposed technique in real life applications.     
			
			{In addition to those primary simulations, we also conducted a few additional studies on two different aspects. In one of those studies, we investigated the performance for the single response case ($J=1$), with only a censored outcome under two treatment options ($K=2$) scenario. Results of this study are provided in the Supplementary Material (see Supplementary Table S1 and S2). Similar to previous results,  we observed increased accuracy in the optimal selection as the training set size increases. In the other study, we investigated the performance of the method under a set of perturbed SIM models (see Supplementary Table S4). Demonstrating the robustness under the departure from SIM structure, results of this study showed comparable performance similar to its counterpart, the Model-2 case (see Supplementary Table S3). }

			\begin{table}[!h]
				\resizebox{1\textwidth}{!}{\begin{tabular}{ccccc}
						\hline
						&  & \multicolumn{3}{c}{Treatment Group $(k)$} \\ \cline{3-5} 
						&  & $k=1$ & $k=2$ & $k=3$ \\ \hline
						\multirow{4}{*}{\begin{tabular}[c]{@{}c@{}}Response \\ $(j)$\end{tabular}} 
						& $j=1$ (cen.) 	& $1+\exp\{ 0.5( 			\bbeta_{11}^{'}X)^3\}   $  &      $1+\exp( -0.5+0.5(-0.8+ \bbeta_{23}^{'}X)^2)	$    &    	$1+\exp\{  1.0-5(     \bbeta_{13}^{'}X)^2\}$  \\ \cline{2-5} 
						& $j=2$ 		& $  \exp\{0.5-2.5(-1.0+  \bbeta_{21}^{'}X)^2\}	$  &      $  \exp(0.5-2.5(1.0  +  \bbeta_{22}^{'}X)^2) 	$    &    	$  \exp(0.5-2.5(        \bbeta_{23}^{'}X)^2) $ \\ \cline{2-5} 
						& $j=3$ 		& $  \exp\{0.5-3(-1.0+    \bbeta_{31}^{'}X)^4\}	$  &      $  \exp(0.5-3 ( 1.0 +   \bbeta_{32}^{'}X)^4) 	$    &    	$  \exp(0.5-3 (         \bbeta_{33}^{'}X)^4) $ \\ \cline{2-5} 
						& $j=4$ 		& $1+\exp(-1.0+  \bbeta_{41}X)						$  & 	  $1+\exp(-1.0-  \bbeta_{42}X)						$  	 & 		$  \exp( 1.0- (\bbeta_{43}X)^2)                $ \\ \hline
				\end{tabular}}
				\caption{Smooth mean functions: Model Set-1. Here, the first response ($j=1$) is considered to be right-censored. Index vectors $\bbeta_{jk}$, $j=1,..,4$, $k=1,...,3$ are provided in Table {\ref{beta_sim}}.}
				\label{SM-1}
			\end{table}

			\begin{table}[!h]
				\resizebox{1\textwidth}{!}{\begin{tabular}{ccccc}
						\hline
						&  & \multicolumn{3}{c}{Treatment Group $(k)$} \\ \cline{3-5} 
						&  & $k=1$ & $k=2$ & $k=3$ \\ \hline
						\multirow{4}{*}{\begin{tabular}[c]{@{}c@{}}Response \\ $(j)$\end{tabular}} 
						& $j=1$ (cen.) 	& $2+\sin\{  \frac{2\pi}{5}+  \pi/2(C^{'}X)\}				$ 	&   $2+\sin\{  \frac{6\pi}{5}+  \frac{\pi}{2}(C^{'}X)\}		$  &  $2+\sin\{ \frac{-2\pi}{5}+  \frac{\pi}{2}(C^{'}X)\}$  \\ \cline{2-5} 
						& $j=2$ 		& $\cos\{   	\frac{\pi}{2}(C^{'}X)\}  						$   &   $\cos\{ \frac{4\pi}{5}+  	\frac{\pi}{2}(C^{'}X)\}  	$  &  $\cos\{\frac{-4\pi}{5}+  	\frac{\pi}{2}(C^{'}X)\}	$  \\ \cline{2-5} 
						& $j=3$ 		& $\sin\{    \frac{\pi}{3}+  	\frac{\pi}{3}(C^{'}X)\}   	$   &   $\sin\{   \pi  +  	\frac{\pi}{3}(C^{'}X)\}   			$  &  $\sin\{\frac{-2\pi}{3}+  	\pi/3(C^{'}X)\} 			$ \\ \cline{2-5} 
						& $j=4$ 		& $\cos\{    \frac{\pi}{3}(C^{'}X)\}   						$   &   $\cos\{ \frac{2\pi}{3}+  	\frac{\pi}{3}(C^{'}X)\}   	$  &  $\cos\{\frac{-2\pi}{3} +  	\frac{\pi}{3}(C^{'}X)\} $  \\ \hline
				\end{tabular}}
				\caption{Smooth mean functions: Model Set-2. Here, the first response ($j=1$) is considered to be right-censored. We choose the common vector ${C}$ to be a unit vector; $C=\big(\frac{1}{\sqrt{r}},...\frac{1}{\sqrt{r}}\big)'$, for all combinations of $j$ and $k$.}	
				\label{SM-2}
			\end{table}

			\begin{table}[!h]
				\centering
				\begin{tabular}{|c|c|c|c|c|c|c|}
					\hline
					\multirow{3}{*}{Group} & \multicolumn{2}{c|}{\multirow{2}{*}{\begin{tabular}[c]{@{}c@{}}Random \\ ($\zeta$)\end{tabular}}} & \multicolumn{4}{c|}{Covariate dependent} \\ \cline{4-7} 
					& \multicolumn{2}{c|}{} & \multicolumn{2}{c|}{25\%} & \multicolumn{2}{c|}{50\%} \\ \cline{2-7} 
					& 25\% & 50\% & $\zeta_{1}$ & $\zeta_{2}$ & $\zeta_{1}$ & $\zeta_{2}$ \\ \hline
					$k=1$ & 0.14 & 0.35 & 0.10 & 0.20 & 0.20 & 0.50 \\ \hline
					$k=2$ & 0.14 & 0.35 & 0.10 & 0.20 & 0.20 & 0.50 \\ \hline
					$k=3$ & 0.12 & 0.32 & 0.07 & 0.18 & 0.18 & 0.47 \\ \hline
				\end{tabular}
				\caption{parameters used for generating censoring times for Model Set-1, with random and covariate dependent censoring settings.}
				\label{cen_1}
			\end{table}

			\begin{table}[!h]
				\centering
				\begin{tabular}{|c|c|c|c|c|c|c|}
					\hline
					\multirow{3}{*}{Group} & \multicolumn{2}{c|}{\multirow{2}{*}{\begin{tabular}[c]{@{}c@{}}Random cen. \\ ( $\zeta$)\end{tabular}}} & \multicolumn{4}{c|}{Covariate dep. cen.} \\ \cline{4-7} 
					& \multicolumn{2}{c|}{} & \multicolumn{2}{c|}{25\%} & \multicolumn{2}{c|}{50\%} \\ \cline{2-7} 
					& 25\% & 50\% & $\zeta_{1}$ & $\zeta_{2}$ & $\zeta_{1}$ & $\zeta_{2}$ \\ \hline
					$k=1$ & 0.11 & 0.26 & 0.05 & 0.20 &  0.15 & 0.40 \\ \hline
					$k=2$ & 0.18 & 0.45 & 0.10 & 0.26 &  0.25 & 0.62 \\ \hline
					$k=3$ & 0.23 & 0.55 & 0.15 & 0.30 &  0.35 & 0.70 \\ \hline
				\end{tabular}
				\caption{parameters used for generating censoring times for Model Set-2, with random and covariate dependent censoring settings.}
				\label{cens_2}
			\end{table}

			\begin{table}[!h]
				\centering
				\resizebox{0.6\textwidth}{!}	
				{\begin{tabular}{lcccccccc}
						\hline
						& \multirow{12}{*}{} & \multirow{3}{*}{} 
						& $\bm{\bbeta_{11}}$ & 0.74	& -0.37	& -0.37	& 0.37	& -0.19		\\ \cline{4-9} 
						&  &  & $\bm{\bbeta_{12}}$ & 0.80	& 0.00	& 0.20	& -0.40	& -0.40		\\ \cline{4-9} 
						&  &  & $\bm{\bbeta_{13}}$ & 0.07	& 0.15	& 0.30	& 0.45	& -0.82		\\ \cline{4-9} 
						&  & \multirow{3}{*}{} & 
						$\bm{\bbeta_{21}}$ & 0.23	& 0.15	& 0.45	& -0.83	& -0.15		\\ \cline{4-9} 
						&  &  & $\bm{\bbeta_{22}}$ & -0.69	& 0.51	& 0.34	& 0.34	& -0.17		\\ \cline{4-9} 
						&  &  & $\bm{\bbeta_{23}}$ & 0.63	& 0.21	& 0.32	& -0.53	& -0.42		\\ \cline{4-9} 
						&  & \multirow{3}{*}{} 
						& $\bm{\bbeta_{31}}$ & 0.15	& 0.30	& 0.07	& -0.82	& 0.45		\\ \cline{4-9} 
						&  &  & $\bm{\bbeta_{32}}$ & 0.24	& 0.16	& 0.65	& -0.24	& -0.65		\\ \cline{4-9} 
						&  &  & $\bm{\bbeta_{33}}$ & -0.18	& 0.36	& 0.54	& 0.18	& -0.72		\\ \cline{4-9} 
						&  & \multirow{3}{*}{} 
						& $\bm{\bbeta_{41}}$ & -0.40	& 0.00	& 0.80	& -0.40	& 0.20  	\\ \cline{4-9} 
						&  &  & $\bm{\bbeta_{42}}$ &  0.48	& 0.27	& -0.55	& 0.41	& -0.48		\\ \cline{4-9} 
						&  &  & $\bm{\bbeta_{43}}$ & -0.75	& 0.34	& -0.14	& 0.14	& 0.54 		\\ \hline
				\end{tabular}}
				\caption{Index vectors $\bbeta_{jk}$s, $j=1,...,J$,  $k=1,...,K$, selected for the simulation study with Model Set-1. We specified each $\bm{\bbeta_{jk}^{\prime}}=(\beta_{jk1},...,\beta_{jkr})_{1\times r}$ with $|\bbeta_{jk}|=1$. }
				\label{beta_sim}
			\end{table}
			\clearpage

			% exp 4
			% \usepackage{multirow}
			\begin{table}[]
				{
					\resizebox{1\textwidth}{!}{\begin{tabular}{|c|c|c|c|c|c|c|c|c|c|c|c|c|c|}
							\hline
							\multirow{3}{*}{\begin{tabular}[c]{@{}c@{}}Weights\\ $(w_1,w_2,w_3,w_4)$\end{tabular}} & \multirow{3}{*}{\begin{tabular}[c]{@{}c@{}}Error\\ dist.\end{tabular}} & \multirow{3}{*}{\begin{tabular}[c]{@{}c@{}}Error disp.\\ para.\\ $(\sigma)$\end{tabular}} & \multirow{3}{*}{\begin{tabular}[c]{@{}c@{}}Size\\ $(n)$\end{tabular}} & \multicolumn{5}{c|}{$\rho=0.3$} & \multicolumn{5}{c|}{$\rho=0.7$} \\ \cline{5-14} 
							&  &  &  & \multirow{2}{*}{\begin{tabular}[c]{@{}c@{}}No\\censoring\end{tabular}} & \multicolumn{2}{c|}{Random} & \multicolumn{2}{c|}{Cov. dep.} & \multirow{2}{*}{\begin{tabular}[c]{@{}c@{}}No\\censoring\end{tabular}} & \multicolumn{2}{c|}{Random} & \multicolumn{2}{c|}{Cov. dep.} \\ \cline{6-9} \cline{11-14} 
							&  &  &  &  & 25\% & 50\% & 25\% & 50\% &  & 25\% & 50\% & 25\% & 50\% \\ \hline
							\multirow{18}{*}{(0.7,0.1,0.1, 0.1)} & \multirow{9}{*}{Normal} & \multirow{3}{*}{0.1} 
							& 		100 & 853 & 809 & 760 & 804 & 765 & 838 & 808 & 765 & 801 & 764 \\ \cline{4-14} 
							&  &  & 200 & 867 & 841 & 836 & 840 & 832 & 852 & 830 & 812 & 826 & 810 \\ \cline{4-14} 
							&  &  & 400 & 886 & 860 & 850 & 860 & 846 & 878 & 860 & 850 & 851 & 851 \\ \cline{3-14} 
							&  & \multirow{3}{*}{0.3} & 
							100 & 758 & 707 & 686 & 700 & 683 & 718 & 686 & 619 & 687 & 615 \\ \cline{4-14} 
							&  &  & 200 & 787 & 753 & 735 & 780 & 762 & 757 & 740 & 710 & 755 & 715 \\ \cline{4-14} 
							&  &  & 400 & 794 & 768 & 748 & 765 & 759 & 775 & 772 & 749 & 772 & 769 \\ \cline{3-14} 
							&  & \multirow{3}{*}{0.5} 
							& 		100 & 618 & 584 & 507 & 570 & 501 & 629 & 598 & 501 & 595 & 509 \\ \cline{4-14} 
							&  &  & 200 & 677 & 642 & 611 & 659 & 602 & 664 & 645 & 595 & 621 & 582 \\ \cline{4-14} 
							&  &  & 400 & 690 & 651 & 671 & 676 & 651 & 686 & 642 & 622 & 636 & 625 \\ \cline{2-14} 
							& \multirow{9}{*}{D.E.} & \multirow{3}{*}{0.1} 
							& 		100 & 853 & 812 & 769 & 816 & 762 & 829 & 795 & 751 & 789 & 748 \\ \cline{4-14} 
							&  &  & 200 & 866 & 848 & 823 & 851 & 804 & 857 & 835 & 799 & 825 & 807 \\ \cline{4-14} 
							&  &  & 400 & 886 & 857 & 837 & 859 & 829 & 871 & 842 & 821 & 839 & 833 \\ \cline{3-14} 
							&  & \multirow{3}{*}{0.3} 
							& 		100 & 782 & 720 & 695 & 723 & 690 & 766 & 715 & 689 & 718 & 690 \\ \cline{4-14} 
							&  &  & 200 & 800 & 783 & 744 & 778 & 752 & 779 & 750 & 731 & 756 & 739 \\ \cline{4-14} 
							&  &  & 400 & 815 & 805 & 792 & 806 & 789 & 798 & 772 & 760 & 769 & 760 \\ \cline{3-14} 
							&  & \multirow{3}{*}{0.5} 
							& 		100 & 655 & 607 & 544 & 605 & 525 & 663 & 602 & 520 & 619 & 541 \\ \cline{4-14} 
							&  &  & 200 & 715 & 686 & 621 & 689 & 614 & 684 & 646 & 608 & 676 & 603 \\ \cline{4-14} 
							&  &  & 400 & 735 & 720 & 703 & 715 & 696 & 707 & 690 & 650 & 686 & 648 \\ \hline
							\multicolumn{14}{|c|}{} \\ \hline
							\multirow{18}{*}{(1.0,1.0,1.0,1.0)} & \multirow{9}{*}{Normal} & \multirow{3}{*}{0.1} 
							& 		100 & 872 & 852 & 829 & 856 & 831 & 859 & 837 & 814 & 821 & 812 \\ \cline{4-14} 
							&  &  & 200 & 887 & 875 & 861 & 874 & 864 & 868 & 845 & 837 & 849 & 836 \\ \cline{4-14} 
							&  &  & 400 & 902 & 888 & 879 & 889 & 876 & 879 & 860 & 851 & 861 & 848 \\ \cline{3-14} 
							&  & \multirow{3}{*}{0.3} 
							& 		100 & 831 & 810 & 798 & 807 & 779 & 730 & 709 & 698 & 714 & 692 \\ \cline{4-14} 
							&  &  & 200 & 838 & 827 & 812 & 826 & 804 & 782 & 770 & 747 & 773 & 741 \\ \cline{4-14} 
							&  &  & 400 & 848 & 826 & 818 & 824 & 817 & 794 & 761 & 755 & 761 & 752 \\ \cline{3-14} 
							&  & \multirow{3}{*}{0.5} 
							& 		100 & 697 & 683 & 645 & 689 & 665 & 653 & 630 & 607 & 627 & 606 \\ \cline{4-14} 
							&  &  & 200 & 730 & 715 & 707 & 717 & 708 & 667 & 638 & 633 & 637 & 626 \\ \cline{4-14} 
							&  &  & 400 & 751 & 749 & 742 & 740 & 730 & 679 & 639 & 629 & 635 & 629 \\ \cline{2-14} 
							& \multirow{9}{*}{D.E.} & \multirow{3}{*}{0.1} 
							& 		100 & 894 & 879 & 859 & 872 & 855 & 847 & 829 & 814 & 830 & 814 \\ \cline{4-14} 
							&  &  & 200 & 896 & 889 & 876 & 886 & 874 & 852 & 844 & 838 & 840 & 833 \\ \cline{4-14} 
							&  &  & 400 & 910 & 903 & 891 & 904 & 889 & 877 & 879 & 847 & 879 & 866 \\ \cline{3-14} 
							&  & \multirow{3}{*}{0.3} 
							& 		100 & 843 & 820 & 794 & 812 & 793 & 783 & 767 & 740 & 762 & 741 \\ \cline{4-14} 
							&  &  & 200 & 850 & 822 & 804 & 814 & 804 & 790 & 776 & 770 & 778 & 767 \\ \cline{4-14} 
							&  &  & 400 & 861 & 846 & 836 & 845 & 834 & 807 & 797 & 784 & 798 & 788 \\ \cline{3-14} 
							&  & \multirow{3}{*}{0.5} 
							& 		100 & 746 & 737 & 706 & 735 & 707 & 699 & 672 & 644 & 671 & 645 \\ \cline{4-14} 
							&  &  & 200 & 773 & 759 & 741 & 764 & 740 & 710 & 698 & 677 & 690 & 664 \\ \cline{4-14} 
							&  &  & 400 & 789 & 760 & 754 & 759 & 749 & 722 & 705 & 706 & 719 & 707 \\ \hline
					\end{tabular}}
					\caption{Accuracies of treatment selection in 1000 test cases using the proposed technique for the case of three treatments and four responses with Model Set-1.}}
			\end{table}

			% 3 exp
			% \usepackage{multirow}
			\begin{table}[]
				{
					\resizebox{1\textwidth}{!}{\begin{tabular}{|c|c|c|c|c|c|c|c|c|c|c|c|c|c|}
							\hline
							\multirow{3}{*}{\begin{tabular}[c]{@{}c@{}}Weights\\ {$(w_1,w_2,w_3)$}\end{tabular}} & \multirow{3}{*}{\begin{tabular}[c]{@{}c@{}}Error\\ dist.\end{tabular}} & \multirow{3}{*}{\begin{tabular}[c]{@{}c@{}}Error disp.\\ para.\\ $(\sigma)$\end{tabular}} & \multirow{3}{*}{\begin{tabular}[c]{@{}c@{}}Size\\ $(n)$\end{tabular}} & \multicolumn{5}{c|}{$\rho=0.3$} & \multicolumn{5}{c|}{$\rho=0.7$} \\ \cline{5-14} 
							&  &  &  & \multirow{2}{*}{\begin{tabular}[c]{@{}c@{}}No\\censoring\end{tabular}} & \multicolumn{2}{c|}{Random} & \multicolumn{2}{c|}{Cov. dep.} & \multirow{2}{*}{\begin{tabular}[c]{@{}c@{}}No\\censoring\end{tabular}} & \multicolumn{2}{c|}{Random} & \multicolumn{2}{c|}{Cov. dep.} \\ \cline{6-9} \cline{11-14} 
							&  &  &  &  & 25\% & 50\% & 25\% & 50\% &  & 25\% & 50\% & 25\% & 50\% \\ \hline
							\multirow{18}{*}{(0.8,0.1,0.1)} & \multirow{9}{*}{Normal} & \multirow{3}{*}{0.1} 
							& 100 		& 844 & 811 & 778 & 809 & 776 & 841 & 803 & 667 & 808 & 669 \\ \cline{4-14} 
							&  &  & 200 & 829 & 826 & 815 & 830 & 808 & 857 & 829 & 802 & 837 & 801 \\ \cline{4-14} 
							&  &  & 400 & 853 & 852 & 833 & 857 & 835 & 874 & 859 & 821 & 855 & 827 \\ \cline{3-14} 
							&  & \multirow{3}{*}{0.3} 
							& 100 		& 722 & 690 & 647 & 686 & 642 & 745 & 736 & 663 & 691 & 649 \\ \cline{4-14} 
							&  &  & 200 & 756 & 720 & 699 & 737 & 704 & 760 & 750 & 696 & 739 & 701 \\ \cline{4-14} 
							&  &  & 400 & 785 & 751 & 740 & 755 & 729 & 780 & 761 & 747 & 761 & 732 \\ \cline{3-14} 
							&  & \multirow{3}{*}{0.5} 
							& 100 		& 619 & 578 & 538 & 580 & 539 & 641 & 612 & 528 & 616 & 526 \\ \cline{4-14} 
							&  &  & 200 & 660 & 624 & 599 & 615 & 602 & 655 & 632 & 588 & 641 & 599 \\ \cline{4-14} 
							&  &  & 400 & 676 & 634 & 625 & 649 & 627 & 679 & 662 & 621 & 672 & 634 \\ \cline{2-14} 
							& \multirow{9}{*}{D.E.} & \multirow{3}{*}{0.1} 
							& 100 		& 843 & 803 & 769 & 801 & 772 & 829 & 798 & 754 & 795 & 766 \\ \cline{4-14} 
							&  &  & 200 & 849 & 820 & 798 & 820 & 808 & 851 & 825 & 801 & 824 & 792 \\ \cline{4-14} 
							&  &  & 400 & 861 & 838 & 822 & 835 & 821 & 864 & 844 & 819 & 837 & 821 \\ \cline{3-14} 
							&  & \multirow{3}{*}{0.3} 
							& 100 		& 757 & 704 & 674 & 701 & 668 & 758 & 701 & 664 & 708 & 663 \\ \cline{4-14} 
							&  &  & 200 & 771 & 761 & 763 & 774 & 765 & 763 & 735 & 699 & 738 & 710 \\ \cline{4-14} 
							&  &  & 400 & 795 & 792 & 773 & 785 & 780 & 789 & 762 & 724 & 763 & 746 \\ \cline{3-14} 
							&  & \multirow{3}{*}{0.5} 
							& 100 		& 659 & 608 & 540 & 609 & 545 & 632 & 596 & 540 & 601 & 542 \\ \cline{4-14} 
							&  &  & 200 & 686 & 677 & 620 & 674 & 619 & 703 & 663 & 619 & 659 & 616 \\ \cline{4-14} 
							&  &  & 400 & 711 & 698 & 671 & 702 & 681 & 718 & 684 & 651 & 688 & 654	 \\ \hline
							\multicolumn{14}{|c|}{} \\ \hline	
							\multirow{18}{*}{(1.0,1.0,1.0)} & \multirow{9}{*}{Normal} & \multirow{3}{*}{0.1} 
							& 100 		& 881 & 838 & 797 & 832 & 800 & 848 & 809 & 782 & 805 & 786 \\ \cline{4-14} 
							&  &  & 200 & 885 & 861 & 845 & 869 & 843 & 865 & 841 & 829 & 843 & 824 \\ \cline{4-14} 
							&  &  & 400 & 898 & 870 & 847 & 872 & 861 & 868 & 865 & 849 & 867 & 845 \\ \cline{3-14} 
							&  & \multirow{3}{*}{0.3} 
							& 100 		& 807 & 764 & 714 & 756 & 720 & 793 & 748 & 707 & 750 & 691 \\ \cline{4-14} 
							&  &  & 200 & 811 & 797 & 773 & 789 & 781 & 801 & 770 & 747 & 767 & 746 \\ \cline{4-14} 
							&  &  & 400 & 842 & 838 & 817 & 838 & 818 & 805 & 777 & 755 & 781 & 756 \\ \cline{3-14} 
							&  & \multirow{3}{*}{0.5} 
							& 100 		& 694 & 663 & 614 & 654 & 618 & 682 & 639 & 607 & 628 & 594 \\ \cline{4-14} 
							&  &  & 200 & 731 & 709 & 671 & 694 & 677 & 696 & 652 & 642 & 667 & 651 \\ \cline{4-14} 
							&  &  & 400 & 747 & 715 & 686 & 713 & 693 & 705 & 707 & 687 & 700 & 685 \\ \cline{2-14} 
							& \multirow{9}{*}{D.E.} & \multirow{3}{*}{0.1} &
							100 		& 881 & 859 & 845 & 857 & 844 & 855 & 828 & 815 & 832 & 817 \\ \cline{4-14} 
							&  &  & 200 & 885 & 868 & 857 & 870 & 861 & 870 & 867 & 833 & 847 & 833 \\ \cline{4-14} 
							&  &  & 400 & 909 & 872 & 857 & 869 & 860 & 884 & 857 & 843 & 852 & 846 \\ \cline{3-14} 
							&  & \multirow{3}{*}{0.3} 
							& 100 		& 826 & 792 & 780 & 797 & 788 & 788 & 762 & 752 & 771 & 756 \\ \cline{4-14} 
							&  &  & 200 & 832 & 819 & 804 & 820 & 817 & 796 & 777 & 761 & 773 & 762 \\ \cline{4-14} 
							&  &  & 400 & 840 & 834 & 828 & 844 & 827 & 818 & 816 & 806 & 817 & 804 \\ \cline{3-14} 
							&  & \multirow{3}{*}{0.5} 
							& 100 		& 715 & 692 & 681 & 685 & 671 & 689 & 650 & 632 & 643 & 631 \\ \cline{4-14} 
							&  &  & 200 & 753 & 741 & 727 & 745 & 729 & 706 & 700 & 692 & 699 & 686 \\ \cline{4-14} 
							&  &  & 400 & 774 & 760 & 741 & 776 & 750 & 732 & 734 & 722 & 730 & 729 \\ \hline
					\end{tabular}}
					\caption{Accuracies of treatment selection in 1000 test cases using the proposed technique for the case of three treatments and three responses with Model Set-1.}}
			\end{table}

			% exp 4
			% \usepackage{multirow}
			\begin{table}[]
				{
					\resizebox{1\textwidth}{!}{\begin{tabular}{|c|c|c|c|c|c|c|c|c|c|c|c|c|c|}
							\hline
							\multirow{3}{*}{\begin{tabular}[c]{@{}c@{}}Weights\\ $(w_1,w_2,w_3,w_4)$\end{tabular}} & \multirow{3}{*}{\begin{tabular}[c]{@{}c@{}}Error\\ dist.\end{tabular}} & \multirow{3}{*}{\begin{tabular}[c]{@{}c@{}}Error disp.\\ para.\\ $(\sigma)$\end{tabular}} & \multirow{3}{*}{\begin{tabular}[c]{@{}c@{}}Size\\ $(n)$\end{tabular}} & \multicolumn{5}{c|}{$\rho=0.3$} & \multicolumn{5}{c|}{$\rho=0.7$} \\ \cline{5-14} 
							&  &  &  & \multirow{2}{*}{\begin{tabular}[c]{@{}c@{}}No\\censoring\end{tabular}} & \multicolumn{2}{c|}{Random} & \multicolumn{2}{c|}{Cov. dep.} & \multirow{2}{*}{\begin{tabular}[c]{@{}c@{}}No\\censoring\end{tabular}} & \multicolumn{2}{c|}{Random} & \multicolumn{2}{c|}{Cov. dep.} \\ \cline{6-9} \cline{11-14} 
							&  &  &  &  & 25\% & 50\% & 25\% & 50\% &  & 25\% & 50\% & 25\% & 50\% \\ \hline
							\multirow{18}{*}{(0.7,0.1,0.1, 0.1)} & \multirow{9}{*}{Normal} & \multirow{3}{*}{0.1} 
							& 		100 & 921 & 890 & 835 & 891 & 834 & 914 & 882 & 832 & 883 & 841 \\ \cline{4-14} 
							&  &  & 200 & 938 & 911 & 895 & 919 & 887 & 924 & 906 & 879 & 906 & 890 \\ \cline{4-14} 
							&  &  & 400 & 946 & 924 & 918 & 933 & 916 & 939 & 917 & 899 & 924 & 909 \\ \cline{3-14} 
							&  & \multirow{3}{*}{0.3} & 
							100 & 857 & 789 & 756 & 786 & 743 & 867 & 793 & 739 & 782 & 727 \\ \cline{4-14} 
							&  &  & 200 & 881 & 821 & 788 & 819 & 897 & 878 & 826 & 785 & 817 & 792 \\ \cline{4-14} 
							&  &  & 400 & 896 & 857 & 817 & 861 & 828 & 891 & 869 & 821 & 870 & 819 \\ \cline{3-14} 
							&  & \multirow{3}{*}{0.5} 
							& 		100 & 799 & 733 & 682 & 741 & 691 & 788 & 716 & 682 & 707 & 674 \\ \cline{4-14} 
							&  &  & 200 & 836 & 798 & 724 & 783 & 722 & 795 & 744 & 711 & 735 & 706 \\ \cline{4-14} 
							&  &  & 400 & 842 & 818 & 763 & 815 & 752 & 812 & 779 & 739 & 774 & 737 \\ \cline{2-14} 
							& \multirow{9}{*}{D.E.} & \multirow{3}{*}{0.1} 
							& 		100 & 934 & 895 & 832 & 889 & 825 & 921 & 887 & 823 & 881 & 815 \\ \cline{4-14} 
							&  &  & 200 & 943 & 914 & 886 & 918 & 879 & 933 & 905 & 880 & 908 & 890 \\ \cline{4-14} 
							&  &  & 400 & 959 & 933 & 905 & 929 & 902 & 945 & 920 & 916 & 926 & 912 \\ \cline{3-14} 
							&  & \multirow{3}{*}{0.3} 
							& 		100 & 862 & 796 & 733 & 803 & 724 & 842 & 790 & 728 & 804 & 729 \\ \cline{4-14} 
							&  &  & 200 & 890 & 837 & 787 & 832 & 891 & 864 & 834 & 780 & 825 & 774 \\ \cline{4-14} 
							&  &  & 400 & 896 & 856 & 833 & 865 & 821 & 878 & 853 & 810 & 856 & 812 \\ \cline{3-14} 
							&  & \multirow{3}{*}{0.5} 
							& 		100 & 811 & 767 & 710 & 754 & 704 & 787 & 707 & 688 & 707 & 671 \\ \cline{4-14} 
							&  &  & 200 & 830 & 798 & 782 & 800 & 769 & 796 & 776 & 720 & 767 & 718 \\ \cline{4-14} 
							&  &  & 400 & 852 & 835 & 816 & 831 & 807 & 822 & 808 & 769 & 801 & 754 \\ \hline
							\multicolumn{14}{|c|}{} \\ \hline
							\multirow{18}{*}{(1.0,1.0,1.0,1.0)} & \multirow{9}{*}{Normal} & \multirow{3}{*}{0.1} 
							& 		100 & 924 & 906 & 896 & 912 & 896 & 926 & 906 & 889 & 906 & 898 \\ \cline{4-14} 
							&  &  & 200 & 930 & 910 & 902 & 918 & 905 & 939 & 917 & 904 & 923 & 907 \\ \cline{4-14} 
							&  &  & 400 & 942 & 929 & 917 & 934 & 917 & 951 & 924 & 916 & 922 & 918 \\ \cline{3-14} 
							&  & \multirow{3}{*}{0.3} 
							& 		100 & 889 & 867 & 855 & 864 & 856 & 864 & 853 & 832 & 851 & 839 \\ \cline{4-14} 
							&  &  & 200 & 896 & 874 & 861 & 877 & 863 & 878 & 869 & 851 & 866 & 857 \\ \cline{4-14} 
							&  &  & 400 & 904 & 878 & 878 & 876 & 869 & 884 & 872 & 857 & 870 & 852 \\ \cline{3-14} 
							&  & \multirow{3}{*}{0.5} 
							& 		100 & 848 & 828 & 805 & 829 & 805 & 821 & 800 & 782 & 792 & 769 \\ \cline{4-14} 
							&  &  & 200 & 858 & 834 & 825 & 835 & 822 & 844 & 824 & 805 & 835 & 802\\ \cline{4-14} 
							&  &  & 400 & 867 & 854 & 842 & 857 & 843 & 861 & 847 & 825 & 850 & 815 \\ \cline{2-14} 
							& \multirow{9}{*}{D.E.} & \multirow{3}{*}{0.1} 
							& 		100 & 951 & 927 & 904 & 921 & 889 & 903 & 881 & 877 & 886 & 876 \\ \cline{4-14} 
							&  &  & 200 & 959 & 921 & 904 & 932 & 909 & 925 & 909 & 898 & 914 & 907 \\ \cline{4-14} 
							&  &  & 400 & 967 & 947 & 935 & 944 & 926 & 938 & 929 & 914 & 921 & 915 \\ \cline{3-14} 
							&  & \multirow{3}{*}{0.3} 
							& 		100 & 893 & 873 & 844 & 873 & 842 & 849 & 836 & 812 & 824 & 811 \\ \cline{4-14} 
							&  &  & 200 & 897 & 866 & 866 & 884 & 866 & 878 & 842 & 832 & 869 & 829 \\ \cline{4-14} 
							&  &  & 400 & 910 & 890 & 870 & 891 & 880 & 854 & 851 & 846 & 852 & 838 \\ \cline{3-14} 
							&  & \multirow{3}{*}{0.5} 
							& 		100 & 834 & 805 & 792 & 816 & 792 & 814 & 788 & 765 & 785 & 763 \\ \cline{4-14} 
							&  &  & 200 & 858 & 842 & 827 & 844 & 827 & 837 & 806 & 795 & 804 & 799 \\ \cline{4-14} 
							&  &  & 400 & 881 & 869 & 852 & 860 & 850 & 853 & 845 & 826 & 832 & 826 \\ \hline
					\end{tabular}}
					\caption{Accuracies of treatment selection in 1000 test cases using the proposed technique for the case of three treatments and four responses with Model Set-2.}}
			\end{table}
			
			% 3 sin
			% \usepackage{multirow}
			\begin{table}[]
				{
					\resizebox{1\textwidth}{!}{\begin{tabular}{|c|c|c|c|c|c|c|c|c|c|c|c|c|c|}
							\hline
							\multirow{3}{*}{\begin{tabular}[c]{@{}c@{}}Weights\\ ${(w_1,w_2,w_3)}$\end{tabular}} & \multirow{3}{*}{\begin{tabular}[c]{@{}c@{}}Error\\ dist.\end{tabular}} & \multirow{3}{*}{\begin{tabular}[c]{@{}c@{}}Error disp.\\ para.\\ $(\sigma)$\end{tabular}} & \multirow{3}{*}{\begin{tabular}[c]{@{}c@{}}Size\\ $(n)$\end{tabular}} & \multicolumn{5}{c|}{$\rho=0.3$} & \multicolumn{5}{c|}{$\rho=0.7$} \\ \cline{5-14} 
							&  &  &  & \multirow{2}{*}{\begin{tabular}[c]{@{}c@{}}No\\censoring\end{tabular}} & \multicolumn{2}{c|}{Random} & \multicolumn{2}{c|}{Cov. dep.} & \multirow{2}{*}{\begin{tabular}[c]{@{}c@{}}No\\censoring\end{tabular}} & \multicolumn{2}{c|}{Random} & \multicolumn{2}{c|}{Cov. dep.} \\ \cline{6-9} \cline{11-14} 
							&  &  &  &  & 25\% & 50\% & 25\% & 50\% &  & 25\% & 50\% & 25\% & 50\% \\ \hline
							\multirow{18}{*}{(0.8,0.1,0.1)} & \multirow{9}{*}{Normal} & \multirow{3}{*}{0.1} 
							& 		100 & 926 & 862 & 826 & 858 & 818 & 927 & 880 & 818 & 865 & 816 \\ \cline{4-14} 
							&  &  & 200 & 937 & 901 & 867 & 900 & 851 & 937 & 905 & 885 & 901 & 877 \\ \cline{4-14} 
							&  &  & 400 & 947 & 917 & 899 & 916 & 897 & 946 & 919 & 904 & 923 & 897 \\ \cline{3-14} 
							&  & \multirow{3}{*}{0.3} 
							& 		100 & 844 & 799 & 733 & 786 & 721 & 854 & 780 & 717 & 775 & 712 \\ \cline{4-14} 
							&  &  & 200 & 873 & 834 & 789 & 810 & 785 & 876 & 829 & 791 & 821 & 780 \\ \cline{4-14} 
							&  &  & 400 & 890 & 860 & 830 & 854 & 806 & 897 & 859 & 816 & 861 & 818 \\ \cline{3-14} 
							&  & \multirow{3}{*}{0.5} 
							& 		100 & 774 & 706 & 667 & 708 & 657 & 780 & 708 & 663 & 709 & 650 \\ \cline{4-14} 
							&  &  & 200 & 791 & 739 & 709 & 732 & 702 & 790 & 753 & 702 & 747 & 702 \\ \cline{4-14} 
							&  &  & 400 & 832 & 801 & 756 & 802 & 749 & 802 & 788 & 735 & 788 & 741 \\ \cline{2-14} 
							& \multirow{9}{*}{D.E.} & \multirow{3}{*}{0.1} 
							& 		100 & 908 & 856 & 807 & 862 & 805 & 923 & 855 & 808 & 857 & 796 \\ \cline{4-14} 
							&  &  & 200 & 922 & 889 & 868 & 903 & 863 & 935 & 902 & 861 & 902 & 858 \\ \cline{4-14} 
							&  &  & 400 & 939 & 915 & 898 & 920 & 894 & 945 & 931 & 903 & 924 & 889 \\ \cline{3-14} 
							&  & \multirow{3}{*}{0.3} 
							& 		100 & 861 & 784 & 731 & 785 & 734 & 869 & 787 & 737 & 781 & 729 \\ \cline{4-14} 
							&  &  & 200 & 883 & 820 & 781 & 818 & 772 & 842 & 814 & 787 & 818 & 783 \\ \cline{4-14} 
							&  &  & 400 & 891 & 859 & 810 & 847 & 812 & 860 & 827 & 815 & 830 & 811 \\ \cline{3-14} 
							&  & \multirow{3}{*}{0.5} 
							& 		100 & 799 & 702 & 661 & 706 & 661 & 817 & 702 & 679 & 716 & 683 \\ \cline{4-14} 
							&  &  & 200 & 816 & 745 & 706 & 746 & 705 & 787 & 738 & 708 & 757 & 707 \\ \cline{4-14} 
							&  &  & 400 & 832 & 798 & 747 & 804 & 758 & 806 & 785 & 742 & 789 & 727 \\ \hline
							\multicolumn{14}{|c|}{} \\ \hline
							\multirow{18}{*}{{(1.0,1.0,1.0)}} & \multirow{9}{*}{Normal} & \multirow{3}{*}{0.1} 
							& 		100 & 929 & 905 & 887 & 905 & 885 & 932 & 917 & 881 & 915 & 893 \\ \cline{4-14} 
							&  &  & 200 & 930 & 916 & 900 & 912 & 908 & 942 & 929 & 904 & 926 & 910 \\ \cline{4-14} 
							&  &  & 400 & 941 & 922 & 907 & 914 & 911 & 957 & 928 & 917 & 928 & 919 \\ \cline{3-14} 
							&  & \multirow{3}{*}{0.3} 
							& 		100 & 869 & 847 & 814 & 843 & 809 & 867 & 855 & 821 & 852 & 822 \\ \cline{4-14} 
							&  &  & 200 & 891 & 879 & 855 & 879 & 846 & 878 & 851 & 847 & 862 & 845 \\ \cline{4-14} 
							&  &  & 400 & 902 & 882 & 869 & 878 & 871 & 894 & 865 & 850 & 869 & 856 \\ \cline{3-14} 
							&  & \multirow{3}{*}{0.5} 
							& 		100 & 830 & 799 & 785 & 805 & 789 & 799 & 772 & 759 & 769 & 751 \\ \cline{4-14} 
							&  &  & 200 & 839 & 819 & 807 & 823 & 812 & 810 & 783 & 779 & 787 & 770 \\ \cline{4-14} 
							&  &  & 400 & 849 & 825 & 825 & 832 & 830 & 821 & 793 & 788 & 793 & 774 \\ \cline{2-14} 
							& \multirow{9}{*}{D.E.} & \multirow{3}{*}{0.1} 
							& 		100 & 924 & 909 & 886 & 905 & 887 & 920 & 904 & 873 & 892 & 878 \\ \cline{4-14} 
							&  &  & 200 & 936 & 916 & 905 & 916 & 901 & 931 & 912 & 897 & 907 & 900 \\ \cline{4-14} 
							&  &  & 400 & 940 & 917 & 911 & 918 & 911 & 949 & 927 & 910 & 928 & 917 \\ \cline{3-14} 
							&  & \multirow{3}{*}{0.3} 
							& 		100 & 889 & 860 & 845 & 858 & 843 & 864 & 848 & 819 & 847 & 818 \\ \cline{4-14} 
							&  &  & 200 & 898 & 879 & 854 & 874 & 860 & 871 & 837 & 828 & 838 & 846 \\ \cline{4-14} 
							&  &  & 400 & 905 & 899 & 887 & 895 & 873 & 887 & 857 & 853 & 857 & 850 \\ \cline{3-14} 
							&  & \multirow{3}{*}{0.5} 
							& 		100 & 836 & 810 & 785 & 812 & 789 & 812 & 789 & 770 & 793 & 763 \\ \cline{4-14} 
							&  &  & 200 & 840 & 823 & 805 & 820 & 810 & 824 & 799 & 790 & 807 & 795 \\ \cline{4-14} 
							&  &  & 400 & 861 & 859 & 842 & 857 & 848 & 830 & 823 & 815 & 824 & 812 \\ \hline			
					\end{tabular}}
					\caption{Accuracies of treatment selection in 1000 test cases using the proposed technique for the case of three treatments and three responses with Model Set-2.}}
			\end{table}
			
			\clearpage
			\section{Illustration using data from the ACTG-175 HIV Randomized Controlled Trial}
			
			In this section, we provide an illustration of our proposed method using, the data resulted from the ACTG 175 Clinical Trial.{[23]} This clinical trial was a randomized, double-blinded, placebo-controlled clinical trial that was conducted to compare single nucleoside or two nucleosides antiviral medications in adults infected with human immunodeficiency (HIV-1) whose T-cell CD4 counts were in the range of 200 to 500 per cubic millimeter. The study randomized HIV-1 infected patients to one of four daily regimens: 600 mg of zidovudine  (arm-0), 600 mg of zidovudine plus 400 mg of didanosine (arm-1), 600 mg of zidovudine plus 2.25 mg of zalcitabine  (arm-2), or 400 mg of didanosine (arm-3). The data set contains information on 2,136 HIV-1–infected subjects. Arms 0, 1, 2, and 3 contain 532, 519, 524, and 561 patients, respectively.{[24]} The primary end-point of this trial was a survival event which we describe in the sequel.  
			
			Since the original treatment assignment in this study was based on random selection, there was no consideration for the optimality. Our intention in this data analysis is simply demonstrate what would be the optimal treatment for a new subject based on his/her characteristics, if training data from an RCT trial were available in advance. In this trial, study participants were periodically examined to capture T-cells counts (i.e., CD4 T helper cells and CD8 cytotoxic T cells), that are critical components in the human immune system. Between these two types of T-cells, CD4 helper cells considered to be the most important component on HIV/AIDS immune response as it involves in suppressing the HIV cell replication via signaling through various other cell types. The main role of the CD8 cell is typically referred to as the antibody reaction against cancers and various types of other viruses. However, studies suggest the important role of the CD8 during the early stages of HIV progression.{[25]} 
			
			An individual's survival endpoint in this study was defined based on three types of events: (i) individual's CD4 count dropping less than 50\% of the pre-treatment count; (ii) an event indicating progression of AIDS; (iii) death. Thus, the term ``survival time" would denote an event time in that sense, that was subjected to right-censoring.
			
			In previous studies, Sriwardhana et al. [5, 9] have been used this dataset to demonstrate personalized treatment strategies with respect to single and dual out-comes. However, the outcomes used in those analyses were completely observed (i.e., uncensored). In the current work, we considered three outcome types, including right-censored survival time, and two uncensored clinical parameters given by CD4 and CD8 counts of a patient observed at week 20. All these outcomes were transformed via log transformation. As covariates, we used log-CD4 and log-CD8 counts at baseline, age, weight, and the number of months a patient received the pre-antiviral therapy.
			
			In our analysis, we selected $200$  patients at random from each treatment arm and considered their trial outcomes and baseline covariates to produce a training dataset. Based on this training data, we estimated the corresponding SIMs for three different outcome types, log-survival (log-T), log-CD4, and log-CD8, coupled with the above covariates. We utilized the IPCW re-weighted SIM for the right-censored log-T outcome. Next, assuming that the remaining 1,336 cases are ``new" patients, we applied the proposed treatment selection for those cases using combinations of weights ranging from 0 to 1 for each response.}
		
		We summarized the resulted assignments for test patients in {Table \ref{ACTG_ASSI}}. {For example, when we chose $\log\mbox{-T}$:$\log \mbox{-CD4}$:$\log\mbox{-CD8}$ weights to be 0.7:0.2:0.1, using the proposed treatment selection, 16,   602,   352, and 366 test patients are proposed to be assigned to arms 0, 1, 2, and 3, respectively; whereas the corresponding assignment is 0, 544, 331, and  461 for corresponding weights 0.3:0.6:0.1.} The overall pattern of these assignments indicates that only a few patients are proposed to be assigned to zidovudine alone, which used as the control arm in the study (arm-0). The marginal analysis of ACTG-175 data concludes, the treatment with Zidovudine and Didanosine combination (arm-1), Zidovudine and Zalcitabine combination (arm-2), or Didanosine alone (arm-3) slows the progression of HIV disease and are superior to treatment with Zidovudine alone.{[23]}  Although the proposed assignment in our approach was based on the patient-specific features, the overall assignment agrees with the above observation. 
		
		\begin{table}[]
			\centering
			{\resizebox{0.75\textwidth}{!}{\begin{tabular}{ccclcccc}
						\hline
						\multicolumn{3}{c}{Weights} &  & \multicolumn{4}{c}{Assignments} \\ \cline{1-3} \cline{5-8} 
						$\omega_{\log\mbox{-T}}$ & $\omega_{\log \mbox{-CD4}}$ & $\omega_{\log\mbox{-CD8}}$ &  & Arm-0 & Arm-1 & Arm-2 & Arm-3 \\ \hline
						0.33 &	0.33 &	0.33 & &    16 (1.20\%) &  602 (45.06\%) &   352 (26.35\%) &   366 (27.40\%) \\ \hline
						0.70 &	0.20 &	0.10 & &	87 (6.51\%) &  258 (19.31\%) &   640 (47.90\%) &   351 (26.27\%) \\ \hline
						0.50 &	0.40 &	0.10 & &	43 (3.22\%) &   366 (27.40\%) &  533 (39.90\%) &   394 (29.49\%) \\ \hline
						0.10 &	0.60 &	0.30 & &	0 (0\%) &  544 (40.72\%) &   331 (24.78\%) &   461 (34.51\%) \\ \hline
			\end{tabular}}}
			\caption{Treatment assignment summary for ACTG-175 Clinical Trial data, by the proposed method based on three types of outcomes: log transformed survival time (${\log\mbox{-T}}$), and two clinical parameters CD4 (${\log \mbox{-CD4}}$) and CD8 (${\log\mbox{-CD8}}$) counts at week 20, using weights $w_{\log\mbox{-T}}$, $w_{\log \mbox{-CD4}}$, and $w_{\log\mbox{-CD8}}$  for CD4 and CD8 counts, respectively. All outcomes were log transformed for the analysis. }
			\label{ACTG_ASSI}
		\end{table}

		%\clearpage
		\section{Discussion}
		
		In this article, we extend the personalized treatment plan suggested by Siriwardhana et al. [9] for a situation where the response data are subject to right-censoring. The method by Siriwardhana et al. [9] intends to select the optimal treatment among multiple treatment options $(K \geq 2)$ when the outcome is multivariate. However, this method uses only the complete data. Our empirical studies show that the modified method performs very satisfactorily in selecting the optimal treatment in a multiple treatment setting and with different censoring scenarios. We demonstrated the proposed method using ACTG-175 trial data analysis, taking three different response variables, including the right-censored survival outcome as one of the outcomes.    
		
		We introduce an IPCW re-weighting scheme to adjust estimators used for the treatment selection and SIM estimation to handle right-censored data. The estimation of IPCW weights is performed using the Aalen’s nonparametric additive model that gives the flexibility of using time-varying covariates. The proposed technique is based on semiparametric SIM models that add great flexibility in modeling real-life situations. The proposed method could also be applied using quantile regression SIMs providing additional model flexibility, compared with the methods based on conditional expectations. However, we did not investigate this option in current research. The problem of selecting the scores summarizing the patient covariates could be an important one. Indeed, we have selected the scores mostly on intuitive grounds. More research on this issue needs to take place.
		
		In the context of personalized medicine, there have been several previous works focused on the optimization of conditional restricted mean survival time given individualized details {(for example see, Tian et al. [26]; Geng et al. [27]; Jiang et al. [28]; Cui et al. [29])}. However, the main goal of these studies was to select the optimal therapy only based on the survival outcome, but not based on multiple responses. 
		
		{Similar to our work, there have been many previous studies utilized the data set from ACTG-175 trial in  various aspects, for illustrating statistical techniques developed on the optimal treatment selection. For example, following a semiparametric concept, {Tsiatis et al.[30]} analyzed the differences in mean CD4 count at week 20, to demonstrate covariate adjustments in RCT data when estimating the mean treatment effects. In their analysis, arm 0 (zidovudine alone ) was compared with combined arms 1-3 (arm-1: zidovudine + didanosine; arm 2 : zidovudine + zalcitabine;  arm 4: didanosine alone).  In a similar analysis setting, Tan et al.{[31]}  used a two-sample empirical likelihood weighting approach to estimate average treatment effect in 20-week CD4 count, while accounting for baseline covariates. In another study, Liang et al.{[32]}, used data-adaptive empirical likelihood-based approach under high-dimensional setting, adopting penalized regression to model the covariate-outcome relationship. They considered many forms of baseline variables, including linear and quadratic forms together with interactions to produce a high dimensional baseline predictor set. In summary, these three studies suggested the marginal superiority of the combined arm (i.e., arms 1-3), compared to arm-0 in terms of 20-week CD4 count response. To improve the efficiency of the log-rank test, Lu and Tsiatis {[33]}  proposed a semi-parametric approach to recover information from auxiliary covariates that are correlating with the survival time. While using ACTG-175 data with multiple predictors of the survival time, they concluded marginal superiority in arm 1, compared to arm 0.  Jiang et al.{[34]} illustrated a nonparametric personalized treatment strategy that maximizes the t-year survival probability with this dataset. Using  arm 0 and arm 1 data, their analysis suggested large assignment proportions to arm 1, in the cases of $t = 600$ days and $t = 800$ days cases. Also, Cui et al.{[35]} analyzed the ACTG-175 survival data to demonstrate a method that based on causal survival forests assess heterogeneous treatment effects. Among many findings, their analysis suggested better outcomes from arm 1, compared to arm 3 for elderly patients. Addressing the multi response problem in personalized treatment case, Siriwardhana and Kualsekera {[36]} used ACTG-175 data considering two outcomes : CD4 and CD8 at week-20. They showed 2\%, 75\%, 11\%, and 12\% assignment rates for arms 0-3, while equally prioritizing two responses. Their analysis showed changes in group allocation rates under different response weights. Similar to the proposed method, Siriwardhana and Kualsekera {[36]} method uses a semiparametric framework together with conditional means, however their method is limited only for fully observed (uncensored data).}  
		
		The proposed methodology for handling multiple responses through rank aggregation is extremely general. Clearly, one could posit different regression models and estimation techniques for different components of the response vector. In addition, different optimality criteria can be used for ranking the treatments for different components of the response vector. The choices of weights are subjective selections based on patient preferences and advice from clinical experts or consultation from suitable support groups.
		{We believe the response weight specification is more of a subjective matter than a quantitative issue where one can guide the choice of weights based on discussions with the patient and advocates, together with advice by clinicians. For responses that are related to the recovery or the progression of the disease, a group of clinical experts may be better suited for deciding the importance of each response in selecting a treatment. Whereas in cases where the outcome measures are related to indicators such as quality of life, economic impact, patient behavior, etc., a clinician with consultation from various support groups may be better suited to determine what weights should be used. While it is useful to find data-driven guidelines, we have not explored those aspects in here. Also, In the current work, we do not objectively account for outcome associations, which is one of the limitations of this work. Even though prioritizing outcomes based on weights could provide a basic solution to the multiple response problem, it could be useful to investigate strategies for prioritizing outcomes while accounting for the natural correlation among responses. It is a worthwhile endeavor to find data-driven guidelines on the weight specification, in future research.}
		
		\clearpage
		%\appendix
		
		\subsubsection*{Compliance with ethical standards}
		Conflict of interest The authors declare that they have no conflict of interest.
		
		\subsubsection*{Acknowledgments}
		The research work by Chathura Siriwardhana was partially supported by grant U54MD007601
		from the National Institutes of Health. Data sharing is not applicable to this article as
		no new data were created or analyzed in this study.

		

		

		\section*{References}
		\begin{enumerate}{}{\setlength{\labelsep}{0in} \setlength{\labelwidth}{0in}
				\setlength{\leftmargin}{.250in} \setlength{\itemindent}{-.260in}}
			
			\item Murphy, S.A. (2003), Optimal dynamic treatment regimes. Journal of the Royal Statistical Society: Series B (Statistical Methodology), 65: 331-355. doi:10.1111/1467-9868.00389
			
			\item van’t Veer, L.J. \& Bernards, R. (2008). Enabling Personalized Cancer Medicine Through Analysis of Gene-Expression Patterns. Nature, 452: 564-570.
			
			\item Vazquez, A. (2013). Optimization of Personalized Therapies for Anticancer Treatment. BMC Syst. Biol., 7: doi:10.1186/1752-0509-7-31.
			
			\item Kosorok, M.R. and Moodie, E.E. (2015). Adaptive Treatment Strategies in Practice: Planning Trials and Analyzing Data for Personalized Medicine. Society for Industrial and Applied Mathematics, doi: 10.1137/1.9781611974188.
			
			\item Siriwardhana, C., Zhao, M., Datta, S., \& Kulasekera, K. (2019). A probability based method for selecting the optimal personalized treatment from multiple treatments. Statistical Methods in Medical Research, 28(3), 749-760.
			
			\item Unnikrishnan AG, Bhattacharyya A, Baruah MP, Sinha B, Dharmalingam M, Rao PV. (2013) Importance of achieving the composite endpoints in diabetes. Indian J Endocrinol Metab, 17(5):835-843. doi:10.4103/2230-8210.117225
			
			\item Johnson P, Greiner W, Al-Dakkak, et al. (2015) Which Metrics Are Appropriate to Describe the Value of New Cancer Therapies? BioMed Research International, https://doi.org/10.1155/2015/865101
			
			\item Fleming TR, Powers JH. Biomarkers and surrogate endpoints in clinical trials. (2012) Stat Med, 31(25):2973-2984. doi:10.1002/sim.5403
			
			\item Siriwardhana, C., Datta, S., Kulasekera, K.B. (2020) Selection of the optimal personalized treatment from multiple treatments with multivariate outcome measures, Journal of Biopharmaceutical Statistics, 30 :(3), 462-480.  DOI: 10.1080/10543406.2019.1684304
			
			\item {Cai T, Tian L, Wong PH, et al. (2011) Analysis of Randomized Comparative Clinical Trial Data for Personalized Treatment Selections. Biostatistics; 12: 270-282.}
			
			\item Pihur, V., Datta, S., \& Datta, S. (2007). Weighted rank aggregation of cluster validation measures: a Monte Carlo cross-entropy approach. Bioinformatics, 23: 1607-1615.
			
			\item Pihur, V., Datta, S., \& Datta, S. (2009). RankAggreg, an R package for weighted rank aggregation. BMC Bioinformatics, 10: 62.
			
			\item Wand, M.P. \& Jones, M.C. (1995) Kernel Smoothing. Chapman \& Hall London.
			
			\item Koul H, Susarla V, Ryzin JV. (1981) Regression Analysis with Randomly Right-Censored Data, Ann. Statist. 9 (6), 1276-1288.
			
			\item Satten GA, Datta S. (2001) The Kaplan-Meier Estimator as an Inverse-Probability-of-Censoring Weighted Average. Am Stat. 55 (3): 207-210.
			
			\item Aalen OO (1980) A model for non-parametric regression analysis of counting processes. In: Klonecki W, Kozek A, Rosiski J (eds) Lecture notes on mathematical statistics and probability, vol 2. Springer, New York, pp 1-25
			
			\item Aalen OO (1980) A linear regression model for the analysis of lifetimes. Stat Med 8:907-925
			
			\item Siriwardhana, C., Kulasekera, K.B. \& Datta, S. (2018) Flexible semi-parametric regres-sion of state occupational probabilities in a multistate model with right-censored data. Lifetime Data Anal 24, 464-491. https://doi.org/10.1007/s10985-017-9403-6
			
			\item Ichimura, H., Hall, P., \& Hardle, W. (1993). Optimal smoothing in single index models. Ann. Stat., 21: 157-178.
			
			\item Siriwardhana (2016). Semi-parametric methods for personalized treatment selection and multi-state models, Doctoral dissertation, University of Louisville, KY USA. September 29 2020. URL: Link: https://ir.library.louisville.edu/ \\cgi/preview.cgi?article=3393\&context=etd 			
			
			\item Genz, A., Bretz, F., Miwa, T., Mi. X., Leisch F., et al. (2017). mvtnorm: Multivariate Normal \& t Distributions, R package version 1.0-3. September 29 2020. URL: https://cran.r-project.org/web/packages/mvtnorm/.
			
			\item Statisticat, LLC. (2017). LaplacesDemon: Complete Environment for Bayesian Inference. Bayesian-Inference.com; R package version 16.0.1. September 29 2020. URL: https://cran.r-project.org/ web/packages/LaplacesDemon/.
			
			\item Hammer, S.M., Katzenstein, D.A., Hughes M.D., Gundacker, H., Schooley, R.T. et al. (1996). A Trial Comparing Nucleoside Monotherapy with Combination Therapy in HIV-Infected Adults with CD4 Cell Counts from 200 to 500 per Cubic Millimeter. N. Engl. J. Med., 335: 1081-1090.
			
			\item Juraska, M., Gilbert P.B., Lu, X., Zhang, M., Davidianet D., et al. (2017). spe 2trial: Semi parametric efficient estimation for a two-sample treatment effect; R package version 1.0.4 2012. September 29 2017. URL: http://cran.r-project.org/package=spe 2trial.
			
			\item Streeck, H. \& Nixon D.F. (2010). T Cell Immunity in Acute HIV-1 Infection. J Infect Dis., 202: 302-308.
			
			\item Tian, L., Zhao, L. and Wei, L. (2014) Predicting the restricted mean event time with the subject’s baseline covariates in survival analysis. Biostatistics, 15 (2), 222-233.
			
			\item Geng, Y., Zhang, H. H. and Lu, W. (2015) On optimal treatment regimes selection for mean survival time. Statistics in Medicine, 34, 1169-1184.
			
			\item Jiang, R., Lu, W., Song, R., Hudgens, M. G. and Naprvavnik, S. (2017) Doubly robust estimation of optimal treatment regimes for survival data with application to an hiv/aids study. The annals of applied statistics, 11 (3), 1763-1786.
			
			\item Cui, Y., Zhu, R. and Kosorok, M. (2017) Tree based weighted learning for estimating individualized treatment rules with censored data. Electronic Journal of statistics, 11, 3927-3953.
			
			\item {Tsiatis, A.A., Davidian, M., Zhang, M., and Lu, X. (2008), Covariate adjustment for two-sample treatment comparisons in randomized clinical trials: A principled yet flexible approach. Statistics in Medicine, 27(23):4658-77}
			
			\item {Lu X. and Tsiatis A.A. (2008), Improving the efficiency of the logrank test using auxiliary covariates. Biometrika, 95(3), 679-694}
			
			\item {Jiang, R., Lu, W., Song, R. and Davidian, M. (2017), On estimation of optimal treatment regimes for maximizing t-year survival probability. J. R. Stat. Soc. B, 79(4):1165-1185}
			
			\item {Cui, Y., Kosorok, M., Sverdrup, E., Wager, S., and Zhu, R. (2021), Estimating heterogeneous treatment effects with right-censored data via causal survival forests. arXiv:2001.09887}
			
			\item {Tan, Y., Wen, X., Liang, W., and Yan, Y. (2020), Empirical likelihood weighted estimation of average treatment effects. arXiv:2008.12989}
			
			\item {Liang, W., and Yan, Y. (2020), Empirical likelihood inference in randomized controlled trials with high-dimensional covariates. arXiv:2010.01772}
			
			\item {Siriwardhana, C., Kulasekera, K.B. (2020), Personalized treatment plans with multivariate outcomes. Biometrical Journal,  62 (8), 1973-1985 }

		\end{enumerate}

		\clearpage

	\end{document}